# Terahertz topological photonics for on-chip communication

*Yihao Yang[1,2], Yuichiro Yamagami[3], Xiongbin Yu[3],*
*Prakash Pitchappa[1,2], Baile Zhang[1,2],*
*Masayuki Fujita[3,\*], Tadao Nagatsuma[3], and Ranjan Singh[1,2,\*]*

[1]Division of Physics and Applied Physics, School of Physical and Mathematical Sciences, Nanyang Technological University, 21 Nanyang Link, Singapore 637371, Singapore.
[2]Centre for Disruptive Photonic Technologies, The Photonics Institute, Nanyang Technological University, 50 Nanyang Avenue, Singapore 639798, Singapore.
[3]Graduate School of Engineering Science, Osaka University, 1-3 Machikaneyama, Toyonaka, Osaka 560-8531, Japan.
Correspondence to: fujita@ee.es.osaka-u.ac.jp, ranjans@ntu.edu.sg
(Date: April 5th, 2019)

**The computing speeds in modern multi-core processors and big data servers are no longer limited by the on-chip transistor density that doubles every two years following the Moore's law, but are limited by the on-chip data communication between memories and microprocessor cores. Realization of integrated, low-cost, and efficient solutions for high speed, on-chip data communications require terahertz (THz) interconnect waveguides with tremendous significance in future THz technology[1-8] including THz-wave integrated circuits and THz communication. However, conventional approaches to THz waveguiding[4,9-11] suffer from sensitivity to defects and considerable bending losses at sharp bends. Here, building on the recently-discovered topological phase of light[12-14], we experimentally demonstrate robust THz topological valley transport on low-loss, all-silicon chips. We show that the valley-polarized topological kink states exhibit unity transmission over a bulk band gap even after propagating through ten sharp corners. Such states are excellent information carriers due to their robustness, single-mode propagation, and linear dispersion – key properties for next generation THz communications. By leveraging the unique properties of kink states, we demonstrate error-free communication through a highly-twisted domain wall at an unprecedented data rate (~10 Gbit/s) and uncompressed 4K high-definition video transmission. Our work provides the first experimental demonstration of the topological phases of THz wave, which could certainly inspire a plethora of research on different types of topological phases in two and three dimensions.**

Terahertz (THz) waves have frequencies in the range of around 0.1 to 10 THz between microwave and optical waves: a region also known as the 'THz gap' due to the lack of many functional devices in this frequency range[1-3,5-8,15-17]. The THz spectral region offers higher available bandwidth that is imperative for meeting the ever growing demand for higher data rates[6,8]. The THz spectral band has the potential to achieve terabits/s data rates over a kilometre distance, which is highly relevant to wireless communication for enabling the internet of things (IoT) with a network of over trillion communicating devices. A critical step needed to achieve this goal is the realization of integrated, low-cost, and efficient solutions for on-chip THz manipulation. Conventional approaches to THz waveguiding include hollow metallic waveguides[10], metallic transmission-lines[18], photonic crystals[11], metal wires[9], and terahertz fibres[4]. However, such conventional approaches suffer from sensitivity to defects (e.g., fabrication imperfections) and considerable bending losses at sharp bends.

Recent discovery of topological phase of light[12-14,19] has provided possible solutions to the above problem. For example, photonic topological insulators (PTIs)[12-14] that insulating in bulk but 'conducting' at edges, show robust edge transport immune to disorders and sharp bends. Great efforts have been made to investigate spin-Hall[20-24] and valley-Hall PTIs (also known as valley-Hall photonic crystals or VPCs)[25-30] that exhibit time-reversal symmetry and do not require magnetic components or temporal modulations. PTIs have already shown excellent potential for applications in modern optical devices such as reflection-less waveguides[31], topological quantum interfaces[32], topological light sources[33], topological lasers[34-36], topological splitters[23], and robust delay lines[37]. However, at present, experimental studies on PTIs have been largely limited to microwave[21,23,24,26,28,31,38] and optical frequencies[20,29,30,32-35,39].

Here, building on the topological phase of light, we experimentally demonstrate robust terahertz topological valley transport on all-silicon chips. To best of our knowledge, it is the first time that the topological phase of light has been introduced into the THz spectral. The on-chip VPCs show an additional merit of being extremely low-loss, which is very important for future THz integrated circuits. In our experiments, we unambiguously demonstrate that valley-polarized topological kink states are very robust and exhibit near unity transmission over a bulk band gap even after passing through ten sharp turns with zero radius of curvature, which includes five 120° turns and five 60° turns. Topological valley kink states are promising for THz communication because of their robustness, single-mode, and linear-dispersion properties. As a proof of concept, we experimentally show error-free communication through a VPC with a highly-twisted domain wall at a high data in addition to successful transmission of uncompressed 4K high-definition (HD) video.

The THz VPCs are patterned on high resistivity 20-kΩ-cm silicon (relative permittivity 11.7) chips with a thickness of $h = 190$ μm, as depicted in Fig. 1(a). The VPC design follows the well-known graphene-like lattice with a lattice constant $a = 242.5$ μm. Each unit cell comprises an equilateral triangle shaped hole with a side length of $l_1$ and another inverted equilateral triangle shaped hole with a side length of $l_2$. In our study, we focus on transverse-electric-like modes, where electric fields in the symmetry plane lie within the plane. The modes propagate in the $xy$ plane but are confined in the $z$-direction due to total internal reflection[40]. In the presence of inversion symmetry (e.g., $l_1 = l_2 = 0.5a$), our VPC possesses $C_6$ symmetry that leads to a pair of degenerate Dirac points (at $K$ and $K'$ valleys) in the band diagram at 0.33 THz (Fig. 1(b)). Upon breaking the inversion symmetry (e.g., $l_1 = 0.65a$ and $l_2 = 0.35a$), the VPC structure reduces to $C_3$ symmetry resulting in the lifting of degeneracy and the disappearance of Dirac points. More importantly, inversion symmetry breaking opens a band gap (0.32 THz $< f <$ 0.35 THz) near the Dirac frequency (Fig. 1(b)). The distributions of the $z$-oriented magnetic field $|H_z|$ and Poynting power flow of the modes at $K$ ($K'$) valley for both bands are shown in Fig. 1(c). It can be clearly seen that there are four modes that exhibit either left-handed circular polarization (LCP) or right-handed circular polarization (RCP). Also, within the same band, the polarization of a mode at $K'$ is opposite to that of a mode at $K$.

It is well-known that VPCs feature non-zero Berry curvatures localized at $K$ and $K'$ valleys[26,28,29,41]. We calculate Berry curvatures based on a first-principle numerical method. Indeed, we find non-zero Berry curvatures localized around different valleys (Fig. 1(d)). Additionally, within the same band, these Berry curvatures are identical but opposite in sign for $K$ and $K'$ valleys. Therefore, the total Berry curvature of any single band is zero. This is also guaranteed by the preservation of time-reversal symmetry in our VPC.

Integrating Berry curvatures around different valleys, we further find that the valley-Chern numbers are half-integer, *i.e.*, $C_K = 1/2$ & $C_{K'} = -1/2$ for the first band and $C_K = -1/2$ & $C_{K'} = 1/2$ for the second band. Our numerical results are consistent with previous works[26,28,29].

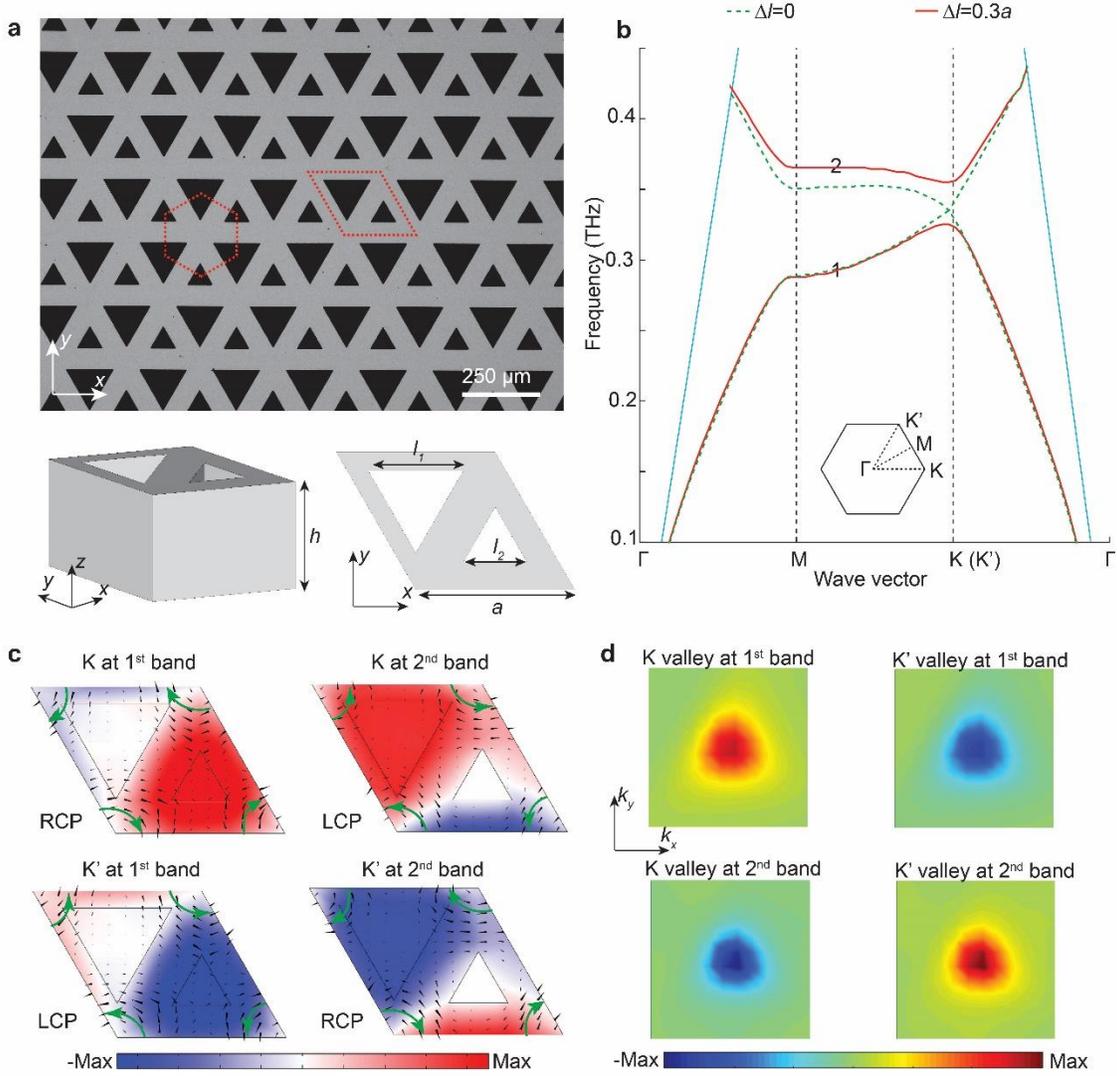

**Figure 1. On-chip THz valley-Hall photonic crystal (VPC) and its bulk band diagram.** (a) An optical image of the fabricated sample. The red dashed lines show Wigner–Seitz and unit cells. Magnified views of the unit cell are presented below the optical image. (b) Band diagrams with and without inversion symmetry. The green, red, and blue lines are the dispersion of the VPC with and without inversion symmetry, and the light line of air, respectively. (c)

Mode profiles at $K$ ($K'$) valleys for the first and second bands of the VPC. The colour labels show the $z$-oriented magnetic field $|H_z|$ while black and green arrows denote the Poynting power flow. (d) Normalized simulated Berry curvatures near $K$ ($K'$) valleys for both bands. The plotted range for each inset is $[-0.33\pi/a, 0.33\pi/a]^2$ centred at the $K$ ($K'$) valley.

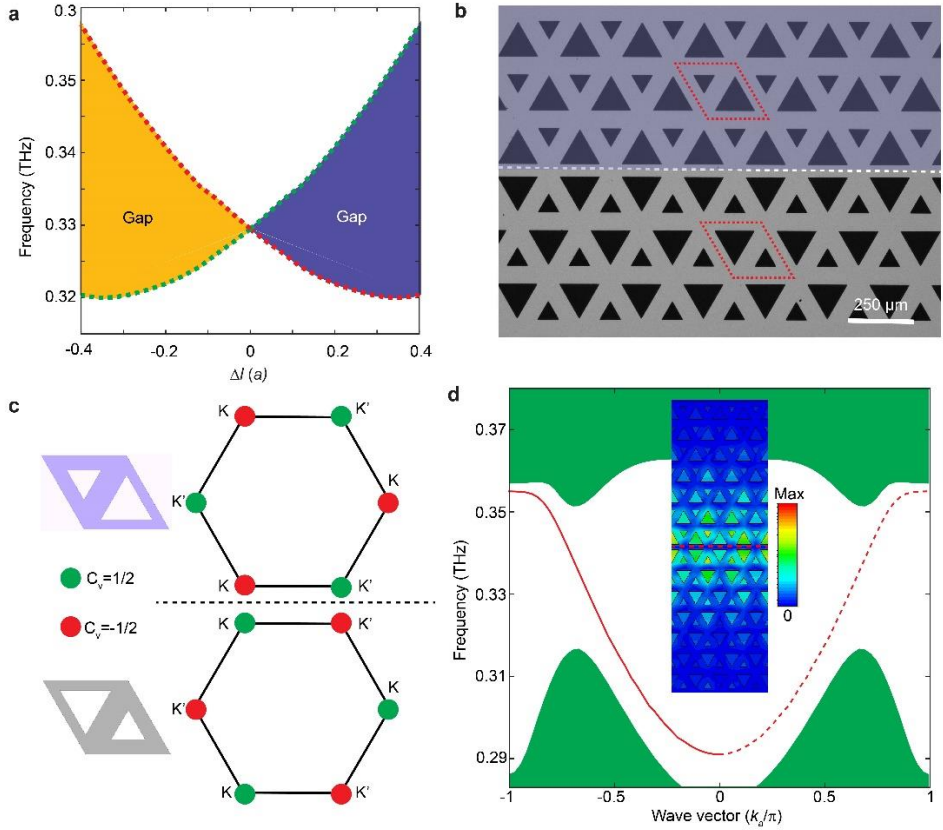

**Figure 2. Phase diagram and topological valley kink states at the domain wall.** (a) A phase diagram showing the variation of the gap as a function of $\Delta l$ in the range of -0.4$a$ to 0.4$a$. The green and red dotted lines represent LCP and RCP, respectively. (b) An optical image of the fabricated domain wall with opposite $\Delta l$ on each side. White dashed line represents the interface between two domains. (c) Topological charge distribution and valley-Chern number on each side of the domain wall. The dashed line denotes the domain wall. (d) Dispersions of topological kink states at the domain wall. The red line and the green regions represent the edge dispersions and the projected bulk dispersions, respectively. The inset shows the intensity distributions of the magnetic field around the domain wall. The red dashed/solid line presents the kink states locked to K/K'.

To visualize the valley-Hall topological phase transition, we numerically obtain the evolution of eigenfrequencies at the *K* valley for both bands as a function of *Δl* (Fig. 2(a)). As shown in Fig. 2a, the obtained results confirm that the polarization of the states is flipped in the phase diagram when *Δl* crosses zero. While LCP is below and RCP is above on the left side (*Δl* < 0) in Fig. 2(a), the polarization flips to LCP above and RCP below on the right side (*Δl* > 0). This inversion indicates that there are two topologically different valley-Hall phases directly related to the signs of *Δl*. Our numerical calculations further show that the valley-Chern numbers for these two valley-Hall phases are exactly opposite.

Next, as shown in Fig. 2(b), we construct a 'kink'-type domain wall between two VPCs with opposite *Δl* values. Difference of valley-Chern numbers across the domain wall at *K* (*K'*) valley is $C_\Delta = \pm 1$ (Fig. 2(c)). According to the bulk-boundary correspondence, a pair of valley-polarized topological kink states appears at the domain wall within the band gap: one locked to the *K* valley propagates forward while the other one locked to the *K'* valley propagates backward. As shown in Fig. 2(d), our simulation results are consistent with this prediction. Interestingly, the dispersions of valley kink states are almost linear in the band gap, which can be explained using the effective 2D massive Dirac Hamiltonian model[41] (see Methods). Furthermore, there is only a single kink state in the band gap for a certain wavevector due to the condition $|C_\Delta| = 1$. We posit that the linear-dispersion and single-mode properties of valley kink states are beneficial for THz communication. The linear-dispersion indicates negligible signal delay at different frequencies, which enables a larger bandwidth[4]. On the other hand, the single-mode feature precludes mode competition and provides additional advantages similar to a single-mode optical fibre[42].

One of the most intriguing properties of topological valley kink states is that they are immune to sharp corners due to the strongly suppressed inter-valley scattering[25,26,28,29,41]. To demonstrate this property, as shown in Fig. 3(a), we fabricate a highly-twisted domain wall on a $26 \times 8$ mm$^2$ silicon chip with ten sharp corners (with a radius of curvature = 0) including five 120° turns and five 60° turns. To experimentally measure the transmission of topological kink states along the twisted domain wall, we set up an experiment as shown in Fig. 3(b). The transmission is measured using an electronic-based continuous wave (CW) THz spectroscopic system in a frequency range between 0.30 to 0.37 THz (see Methods).

As shown in Fig. 3(c), the measured results are compared with the transmission in VPCs comprising either a straight domain wall or no domain wall. A distinct dip in transmission, between 0.32 THz to 0.35 THz, can be clearly seen in a VPC without a domain wall indicating the presence of a bulk band gap. In contrast, the transmission within the band gap is close to unity for VPCs with twisted or straight domain walls. The minimum bending loss is estimated to be < 0.1 dB/bend, which is smaller than that of conventional THz photonic-crystal waveguide with optimization (~0.2 dB/bend)[11]. Although the transmission was similar within the gap, the transmission values for straight and twisted domain walls are different outside the gap due to different scattering mechanisms. We also perform full-wave simulations to visualize the intensity distributions of the magnetic field near the twisted domain wall, as shown in Fig. 3(d). It is evident that a kink state smoothly travels through sharp corners with a negligible reflection. These results evince that the topological kink states are immune to sharp bends in our THz photonic chips.

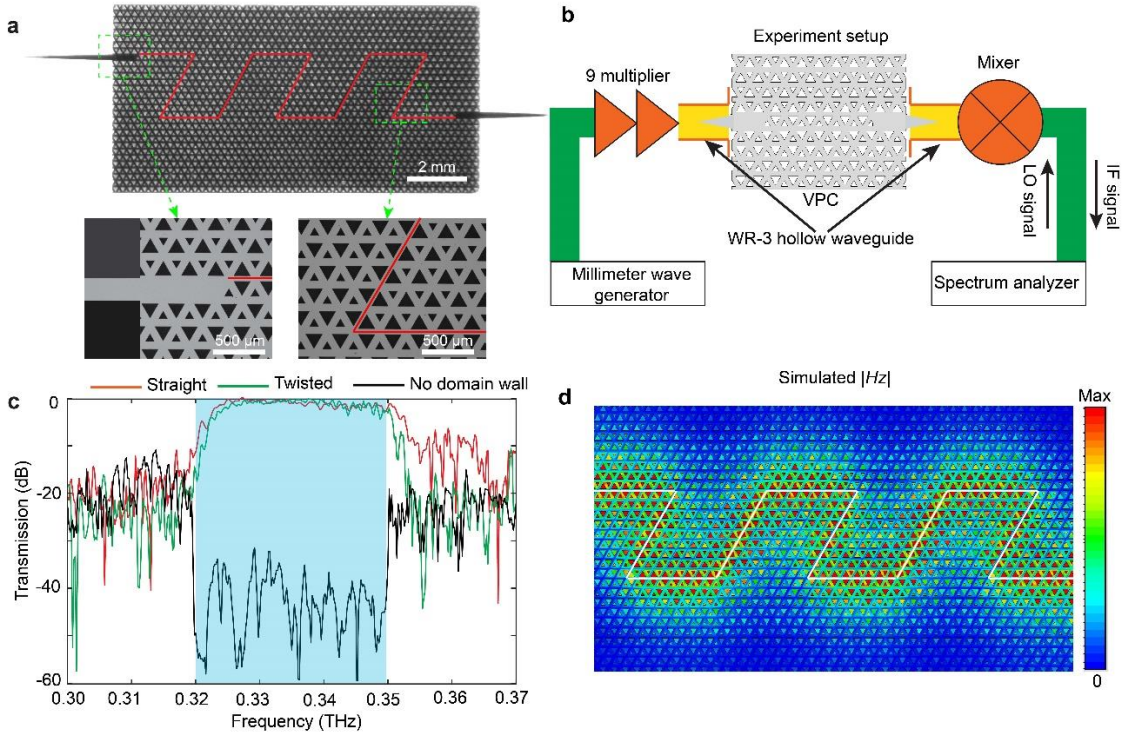

**Figure 3. The experimental observation of robust topological valley kink states along a twisted domain wall in a large-scale THz photonic chip.** (a) An optical image of the fabricated twisted domain wall. The red lines represent

the positions of the domain walls. (b) The experimental setup for measuring transmission. (c) Measured transmission curves for a VPC with a straight domain wall, a twisted domain wall with 10 corners, and no domain wall. The blue region represents the bulk band gap. (d) Simulated $|H_z|$ field distributions in the on-chip VPC at 0.335 THz. The white line denotes the position of the domain wall.

Topological kink states, identified in this article, can be excellent information carriers with applications in on-chip THz communication. To demonstrate this application, we perform a THz communication experiment using the highly-twisted domain wall with ten sharp corners (see Methods). The measured bit error rate as a function of data rate is shown in Fig. 4(a). Clearly, we are able to achieve an error-free (bit error rate $< 10^{-11}$) transmission at a data rate up to 11 Gbit/s with a carrier frequency of 0.335 THz. Remarkably, the achieved bit rate of 11 Gbit/s is higher than that of conventional THz photonic crystal with bends/turns (1.5 Gbits/s)[11]. Additionally, a clear eye diagram obtained on the screen of the oscilloscope at a data rate of 11 Gbit/s is shown in Fig. 4(b).

In order to demonstrate the capability of high-speed and high-quality real-time data transmission, as shown in Fig. 4(c), we set up another experiment for transmitting an uncompressed 4K HD video at a data rate of 6 Gbit/s (see Methods). The transmitted 4K HD video is shown on the monitor in the background. From Fig. 4(c) and a Movie 1 in the Supplementary Information, it can be seen that the 4K video is successfully transmitted in real-time through the VPC even with a twisted domain wall. All the above experiments unambiguously demonstrate that topological valley kink states are robust information carriers for on-chip THz communication.

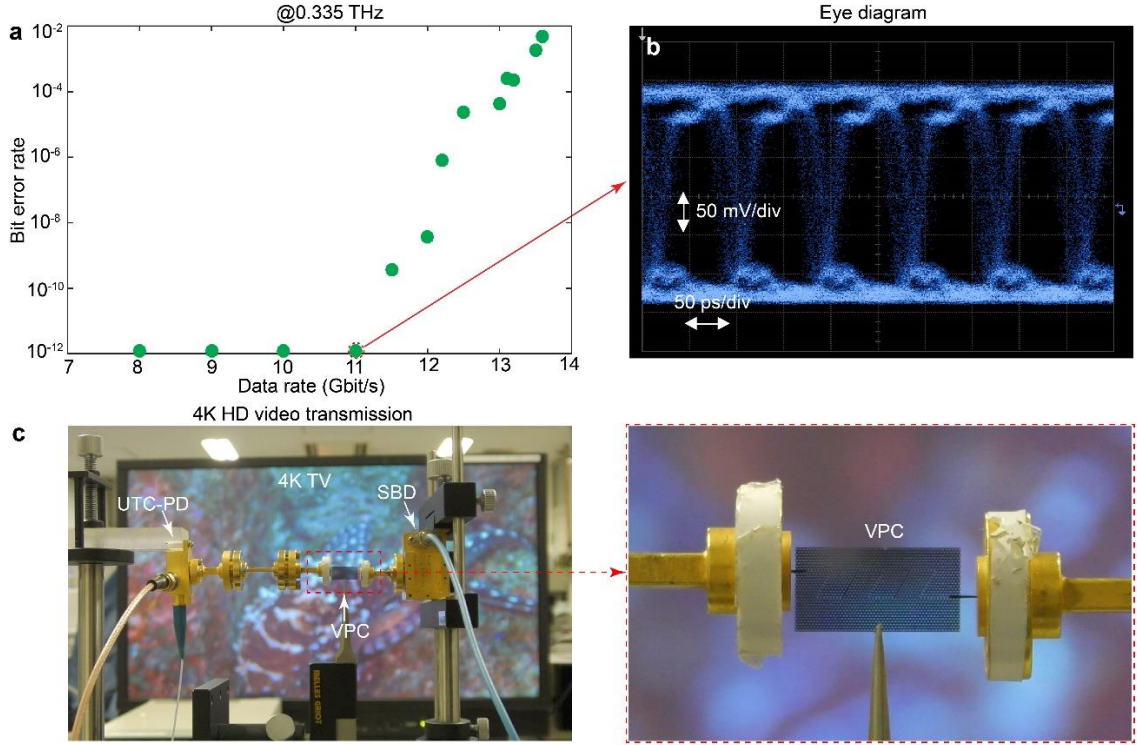

**Figure 4. Terahertz communication based on robust topological valley transport.** (a) The measured bit error rate as a function of the data rate at 0.335 THz. (b) An eye diagram on the oscilloscope at 11 Gbit/s with a bit error rate of $1\times10^{-12}$. (c) An experimental demonstration of uncompressed 4K HD video transmission. The transmitted 4K HD video is shown on the monitor in the background. The dashed red rectangle highlights the VPC with a twisted domain wall. An enlarged picture of the sample is shown on the right.

We thus experimentally demonstrate robust photonic transport on low-loss silicon chips using the valley-Hall topological phase of THz waves. Besides, topological valley kink states are excellent information carriers for THz communication. By combining robust topological transport across sharp corners, single-mode propagation, and linear-dispersion of kink states, we experimentally demonstrate error-free transmission of uncompressed 4K HD video to showcase the great potential of THz topological photonics. Topological kink states are poised to enable futuristic THz functional components such as topological splitters[26], robust delay lines[25], compact interferometers, and directional antennas[28,43] using low-loss and all-silicon VPC chips. Therefore, the present silicon-based VPC would serve as an ideal platform for the next-generation of THz-wave integrated circuits and THz communication. Finally, our work provides the first experimental demonstration of the topological phases of THz wave, which could certainly inspire a plethora of research on different types of topological phases in both two and three dimensions including spin-Hall PTIs[32], quantum-Hall PTIs[31,34], and Weyl points[44-46].

## Acknowledgments

Y. Yang, P. P. and R. S. acknowledge the research funding support from the Singapore Ministry of Education (MOE2017-T2-1-110, and MOE2016-T3-1-006(S)) and the National Research Foundation (NRF), Singapore and Agence Nationale de la Recherche (ANR), France (grant No. NRF2016-NRF-ANR004). Work at Osaka University is supported in part by the Core Research for Evolutional Science & Technology (CREST) program of Japan Science and Technology Agency (#JPMJCR1534), Grant-in-Aid for scientific research from the Ministry of Education, Culture, Sports, Science and Technology of Japan (#17H01764).

## Author Contributions

Y. Yang created the design, performed theoretical analysis, and helped write the manuscript. Y. Yamagami performed simulations, experiments, and data analysis. X. Y. supported the design and performed the experiments. P. P. supported the experiment design and simulation. B. Z. provided inputs on topological protection. M. F. planned and co-led the project, performed the data analysis, and helped write the manuscript. T. N. guided the project and communication experiments. R. S. planned, led the project and helped write the manuscript. All authors contributed to the manuscript.

## Author Information

The authors declare no competing financial interests. Readers are welcome to comment on the online version of the manuscript. Correspondence and requests for materials should be addressed to Drs. Ranjan Singh (ranjans@ntu.edu.sg) and Masayuki Fujita (fujita@ee.es.osaka-u.ac.jp).
12